\documentstyle[preprint,eqsecnum,aps]{revtex}

\newcommand{\be}{\begin{equation}}
\newcommand{\ee}{\end{equation}}
\newcommand{\bea}{\begin{eqnarray}}
\newcommand{\eea}{\end{eqnarray}}
\newcommand{\bml}{\begin{mathletters}}
\newcommand{\eml}{\end{mathletters}}
\newcommand{\nn}{\nonumber}
\newcommand{\sla}{\! \not \!}
\newcommand{\Punkt}{\quad .}
\newcommand{\Komma}{\quad ,}
\newcommand{\otto}{\leftrightarrow}

\begin{document}
\draft

\preprint{SUBATECH--98--16}

\title{Expansion and Hadronization of a Chirally Symmetric
       Quark--Meson Plasma}

\author{P.~Rehberg, L.~Bot and J.~Aichelin}
\address{SUBATECH \\ Laboratoire de Physique Subatomique et des
         Technologies Associ\'ees \\
         UMR Universit\'e de Nantes, IN2P3/CNRS, Ecole des Mines de Nantes \\
         4 Rue Alfred Kastler, F-44070 Nantes Cedex 03, France}

\maketitle
\vspace{2cm}
\begin{abstract}
Using a chirally symmetric Lagrangian, which contains quarks as
elementary degrees of freedom and mesons as bound states, we investigate
the expansion and hadronization of a fireball, which initially contains
only quarks and produces mesons by collisions. For this model, we study
the time scales of expansion and thermal and chemical equilibration.
We find that the expansion progresses relatively fast, leaving not
necessarily enough time to establish thermal and chemical equilibrium.
Mesons are produced in the bulk of the fireball rather than at a surface,
at a temperature below the Mott temperature. Initial density fluctuations
become amplified during the expansion. These observations challenge
the applicability of hydrodynamical approaches to the expansion of a
quark-gluon plasma.
\end{abstract}


\clearpage

\section{Introduction} \label{introsec}
The physics of the quark-gluon plasma (QGP) has formed one of the most
important points of interest in nuclear and high energy physics during
the last years \cite{qm96}. While possible evidence for its creation has
already been obtained in $Pb$\/--$Pb$\/ collisions at the CERN SPS, new
experiments at RHIC and LHC are under construction. It is expected that
these experiments will reproduce the SPS results and beyond that confirm
other signals for the creation of a QGP, which have been proposed in
the literature.

With these improved experiments coming up soon, there is a clear need for
a theoretical description, which is able to describe the different stages
of a heavy ion collision, as there are (i) the formation of the fireball,
(ii) its expansion within the QGP phase, (iii) the phase transition to
the hadronic phase, (iv) the expansion within the hadronic phase, and
(v) the decay of the fireball into noninteracting fragments. With the
exception of the first of these stages, this is frequently done using
hydrodynamical models \cite{hydro} due to the lack of a more realistic
approach. This approach assumes, that one has
local thermal equilibrium during the whole evolution at least for the
light and thus dominant components of the system, so that it is sufficient
to compute the space-time dependence of the energy-momentum tensor using
a phenomenological equation of state, which may be extracted e.\,g. from
lattice simulations. One of the drawbacks of this method is that it is
implicitly assumed that the time scale for collisions is short compared to
the expansion time scale, which is by no means obvious due to the
large expansion velocity of the fireball. Furthermore, the equation
of state is unknown for $\mu=0$, since this sector is still not accessible
for lattice computations. Another weak point is that hydrodynamics has
no intrinsic freezeout mechanism, so that the breakup stage into hadrons
of the expansion must be described by additional assumptions.

Another common approach for the modelization of heavy ion collisions are
transport calculations based on cascade codes. These models have the
advantage of being generic non-equilibrium calculations. They suffer,
however, from the fact that usually the in-medium modifications of
particle properties and interaction cross sections are
neglected. Furthermore, no attempt is made to apply these calculations
to phase transitions. This limits the validity of these models to
the late stages of the expansion, where the system is already in the
hadronic phase.

The approach we present here differs from those presented above in that
we attempt to construct a cascade type model, which is nevertheless able
to describe a transition from a pure quark phase to a hadronic phase,
and which also takes into account the in-medium modifications of quark
and hadron properties. Doing this in a fashion which describes all of
the phenomena is nevertheless impossible, since one of the major effects
of the QGP transition, the confinement, is presently not understood
sufficiently to implement it in a non-equilibrium model. It is, however,
known that confinement does not play an important role in the low energy
hadron phenomenology. This sector is rather dominated by chiral symmetry,
which can be well described by effective models. Since the importance of
chiral symmetry is well established and the behaviour of effective chiral
models is also extensively studied, an incorporation of this symmetry
into the dynamical theory of heavy ion collisions seems mandatory. We
thus take one of these effective models, the Nambu--Jona-Lasinio (NJL)
model \cite{nambu,sandi}, for which a non-equilibrium treatment has been
recently derived \cite{zhawi,wojtek,ogu,trap,pion}, and for which
first numerical solutions have been reported in Refs.~\cite{trap,abada,bot}.
The advantages of this approach are obvious: Since the NJL model starts
from a Lagrangian, which contains only quarks as elementary degrees
of freedom and treats pions as bound states, one is able to describe
the transition of a hot system containing only quarks initially to a
system which contains quarks and hadrons. The technical tools for the
treatment of this transition have been developed in Ref.~\cite{pion}. The
possibility of including both quarks and hadrons, the former appearing
in the initial state while the latter will be present in the final state,
distinguishes the NJL model from most of the other chiral models such as
chiral perturbation theory or the $\sigma$-model \cite{chpt}, while the
description of hadronic properties at low temperatures is equally good
for all of them. A comparison of the NJL mesonic mass spectrum with
the lattice calculations of Ref.~\cite{edwin} shows that also at high
temperatures the NJL model provides a good description of mesonic
properties. Besides that, another advantage of our approach is that it is
possible to derive the whole expansion scenario from one single Lagrange
density. Since the NJL model does, however, not include confinement,
a price has to be paid for this, in that the transition between the
two phases is not given by a confinement transition, but rather by a
Mott effect \cite{gerry}. As a consequence, one has thus free quarks at
all temperatures.

The numerical method we choose for the solution of the transport equations
is an algorithm of the QMD type \cite{qmd}.
In this approach, it is assumed that the particles can be described
by (nonrelativistic) gaussian wave functions, which contain two time
dependent parameters, the position of the centroids in coordinate and
momentum space. The total wavefunction of the system is assumed to be
the product of the wave function of the individual particles. The form of
the wavefunction remains time independent and the time evolution of the
parameters is obtained applying a time dependent variational principle,
which yields time evolution equations very close to the classical time
evolution equations. Being a $n$-body theory, this approach allows
to investigate in detail the time evolutions of correlations and
fluctuations, which is beyond the range of applicability of BUU type
transport theories.  For details we refer the reader to Ref.~\cite{hart}.

Previous approaches to a nonequilibrium treatment of the NJL model
have been given in Refs.~\cite{trap,abada,bot}. This work goes
beyond Refs.~\cite{abada,bot} in that these papers only considered
a system of quarks moving within a mean field, whereas we consider
also collisions. Reference \cite{trap} also included collisions
in a relaxation time approach,
however, mesonic degrees of freedom and the mechanisms of their
creation were not studied there. This is thus the first work, in which
the chiral phase transition is studied including collisions between quarks
as well as a hadronization of quarks into mesons.

This paper is organized as follows: In Sec.~\ref{modelsec}, we
describe our interaction model and its numerical treatment.
Numerical results are given in Sec.~\ref{numsec}. We summarize
and conclude in Sec.~\ref{sumsec}.

\section{Description of the Model and the Algorithm} \label{modelsec}
The model which we use throughout this paper is the Nambu--Jona-Lasinio
(NJL) model \cite{nambu} in its two flavor version, which is defined by the
Lagrangian
\be
{\cal L} = \bar\psi\left( i\sla\partial - m_0\right)\psi
           + G \left[ \left(\bar\psi\psi\right)^2
           + \left(\bar\psi i\gamma_5\vec\tau\psi\right)^2\right]
\Punkt \label{lagra}
\ee
Here, $\psi$ denotes the quark fields, which are implicitly understood
to carry flavor and color indices, and $\vec\tau$ are the Pauli
matrices in flavor space. A current quark mass $m_0$ is
introduced, which provides a small explicit chiral symmetry breaking.

For a review of the equilibrium properties of this model, the reader is
referred to Ref.~\cite{sandi}.  Here we summarize only those points, which
are essential for our calculations. At small temperatures and densities,
the interaction in Eq.~(\ref{lagra}) leads to a spontaneous breakdown
of chiral symmetry, which is described by a finite effective quark mass
$m_q$. To lowest order in an expansion in the inverse number of colors,
$1/N_c$ \cite{expand}, which for quarks is commensurate with the Hartree
approximation, $m_q$ is given by the gap equation
\be
m_q = m_0 + 2Gm_q \int_{|\vec p|<\Lambda} \frac{d^3p}{(2\pi)^3}
      \frac{1}{E_q(\vec p)}
      \left( 2N_cN_f - n_q(\vec p) - n_{\bar q}(\vec p) \right)
\Komma \label{gap}
\ee
where $E_q(\vec p) = \sqrt{\vec p^2 + m_q^2}$ is the quasiparticle energy
of the quarks and $n_q(\vec p)$, $n_{\bar q}(\vec p)$ denote the quark
and antiquark momentum distributions, respectively. Since the NJL model
is non-renormalizable, we have restricted the integral in Eq.~(\ref{gap})
to momenta smaller than a cutoff $\Lambda$.

Pions appear in the NJL model as bound states of quarks and antiquarks
in the pseudoscalar interaction channel. As a consequence of the
Goldstone theorem, they become massless in the chiral limit, $m_0\to
0$. If, however, $m_0$ is finite, they also acquire a finite mass. Their
quasiparticle energy can be extracted from the poles of their effective
retarded propagator, $\Delta^R_\pi(p_0,\vec p)$, which is given by
\cite{sandi}
\be
\Delta^R_\pi(p_0,\vec p) = \frac{2G}{1-2G\Pi^R_{PS}(p_0,\vec p)} \Komma
\ee
where $\Pi^R_{PS}(p_0,\vec p)$ is the irreducible retarded pseudoscalar
polarization function.  The quasiparticle energy of the pion is thus
given as the solution of
\be
1 - 2G\Pi^R_{PS}(E_\pi(\vec p), \vec p) = 0 \Punkt
\label{pidis}
\ee
In the following we will make the approximation $E_\pi(\vec p) =
\sqrt{\vec p^2 + m_\pi^2}$, where $m_\pi$ is computed as $E_\pi(\vec 0)$
from Eq.~(\ref{pidis}). The retarded polarization is computed, again
to lowest order in $1/N_c$, from the diagram given in Fig.~\ref{polarfig}.
In this case, the irreducible pseudoscalar polarization is explicitly given by
\bea
&&\Pi^R_{PS}(p_0, \vec p) = \\ \nn
&&\hspace{1cm} 2 \int_{|\vec p|<\Lambda} \frac{d^3q}{(2\pi)^3}
\frac{2N_cN_f-n_q(\vec q)-n_{\bar q}(\vec q)}{E_q(\vec q)}
\frac{\vec p\vec q(p^2+2\vec p\vec q)-2p_0^2E_q(\vec q)^2}
     {(p^2+2\vec p\vec q)^2-4p_0^2E_q(\vec q)^2+i\epsilon p_0}
\Komma
\eea
with the term $i\epsilon p_0$ in the denominator taking care of the
causal structure.

Due to the scalar coupling channel in the Lagrangian (\ref{lagra}),
the model contains also a scalar resonance, the $\sigma$ meson.
Its mass can be computed from a dispersion relation similar to
Eq.~(\ref{pidis}), replacing the pseudoscalar polarization by
the scalar one,
\bea
&&\Pi^R_{S}(p_0, \vec p) = \\ \nn
&&\hspace{1cm} 2 \int_{|\vec p|<\Lambda} \frac{d^3q}{(2\pi)^3}
\frac{2N_cN_f-n_q(\vec q)-n_{\bar q}(\vec q)}{E_q(\vec q)}
\frac{(\vec p\vec q+ 2m_q^2)(p^2+2\vec p\vec q)-2p_0^2E_q(\vec q)^2}
     {(p^2+2\vec p\vec q)^2-4p_0^2E_q(\vec q)^2+i\epsilon p_0}
\Punkt
\eea
This function is computed from the same graph as given in
Fig.~\ref{polarfig}, with $i\gamma^5\tau$ replaced by one.

The mass spectrum computed from Eqs.~(\ref{gap}), (\ref{pidis}) in thermal
equilibrium is shown in Fig.~\ref{massfig}. The parameters used for
this plot are $m_0=4$\,MeV, $G\Lambda^2=1.989$ and $\Lambda = 820$\,MeV.
At zero temperature, chiral symmetry is spontaneously broken and the
quarks appear as constituent quarks with masses of $m_q=323$\,MeV. As
a consequence, light Goldstone pions appear as bound states, which are,
in the chiral limit, massless, but for the parameters given have the mass
$m_\pi=136$\,MeV. The $\sigma$ meson has a mass of $m_\sigma=653$\,MeV.
As the temperature rises, the Pauli blocking term of Eq.~(\ref{gap})
becomes more and more important. Thus the quark mass drops at high
temperatures and chiral symmetry becomes restored. Concomitantly,
the pion mass will rise, so that at the Mott temperature $T_M$ its
mass becomes equal to that of its constituents, $m_\pi(T_M)=2m_q(T_M)$.
For our parameter set, this transition occurs at $T_M=219$\,MeV. At
temperatures higher than $T_M$, the ground state of the model is no
longer given by the pions, but rather by the quarks, while the pions
themselves become unstable resonances. This instability models, to
a certain extent, the deconfinement transition of QCD. For a more
detailed discussion of this Mott transition the reader is referred
to Ref.~\cite{gerry}. As a consequence of chiral symmetry restauration,
the $\sigma$ mass becomes degenerate with the pion mass at high
temperatures.

It has been detailled in Refs.~\cite{zhawi,wojtek,ogu,pion}, how
the treatment of the NJL model is generalized to non-equilibrium. In
particular, it has been shown, that the quasiparticle energy of the
quarks is given by an equation similar to Eq.~(\ref{gap}), where $m_q$
and the momentum distributions become space-time dependent, whereas
the meson self energy in turn is given by to Eq.~(\ref{pidis}) with a
space-time dependent polarization. The equations of motion for the one
particle distribution functions is then given by
\bml \label{transeq} \bea
\left(\partial_t + \vec\partial_p E_q \vec\partial_x
     - \vec\partial_x E_q \vec\partial_p \right) n_q(t, \vec x, \vec p)
&=& I_{{\rm coll}, q}[n_q, n_\pi] \\
\left(\partial_t + \vec\partial_p E_\pi \vec\partial_x
     - \vec\partial_x E_\pi \vec\partial_p \right) n_\pi(t, \vec x, \vec p)
&=& I_{{\rm coll}, \pi}[n_q, n_\pi] 
\Komma
\eea \eml
where $I_{{\rm coll}, q}[n_q, n_\pi]$ and $I_{{\rm coll}, \pi}[n_q, n_\pi]$
are collision integrals of the Boltzmann type.

In the present paper, we solve Eqs.~(\ref{transeq}) using an QMD-like
approach. To this end, we replace the one particle distribution function
$n_q$ by the parametrization
\be
n_q(t, \vec x, \vec p) = \sum_{i=1}^{N_q}
           \exp\left(-\frac{\left[\vec x - \vec x_i(t)\right]^2}{2w^2}\right)
           \exp\left(-\frac{w^2}{2}\left[\vec p - \vec p_i(t)\right]^2\right)
\label{param}
\ee
and analogously the antiquark distribution function $n_{\bar q}$
and the pion distribution function $n_\pi$. The normalization in
Eq.~(\ref{param}) is chosen in such a way that
\be
\int \frac{d^3x d^3p}{(2\pi)^3} n_q(t, \vec x, \vec p) = N_q
\Komma
\ee
where $N_q$ is the total number of quarks. The center points
$\vec x_i(t)$, $\vec p_i(t)$ in Eq.~(\ref{param}) are moving on the
characteristics of Eq.~(\ref{transeq}):
\bml \label{charact} \bea
\frac{d}{dt} \vec x_i(t) &=& \vec \partial_p E \\
\frac{d}{dt} \vec p_i(t) &=& - \vec \partial_x E
+ \mbox{collision contributions} \label{pmove} \Punkt
\eea \eml
The second term on the right hand side of Eq.~(\ref{pmove}) serves to
describe the effects of the collision integral. It is computed as
follows: for each pair of particles, the coordinates and momenta are
boosted into the respective two particle center of mass frame.
Here, the distance vector of the particles is decomposed into
its longitudinal part, i.\,e. the part parallel to the relative
momentum, and its transverse part, i.\,e. the part perpendicular
to the relative momentum. We decide, which scattering processes are
possible for the pair in question and compute the respective cross
sections $\sigma_k$ as well as the total cross section $\sigma_{\rm
tot}=\sum\sigma_k$.  A collision happens, if (i) the longitudinal distance
of the incoming particles in the CM frame is smaller than the distance
traveled by the particle within the time step and (ii) the transverse
distance is smaller than $\sqrt{\sigma_{\rm tot}/\pi}$. In this case, the
specific process is chosen with probability $\sigma_k/\sigma_{\rm tot}$.
Neglecting the dependence of the differential cross section on the angles,
we choose the direction of the outgoing momenta randomly in the center
of mass system and boost them back into the original reference frame.

The processes implemented so far which are treated in this way belong to
two classes. The first of these are elastic scattering processes of the
form $qq \otto qq$, $q\bar q \otto q\bar q$ and $\bar q\bar q\otto\bar
q\bar q$ \cite{elaste}. The generic Feynman diagrams of these processes
in leading order of the $1/N_c$ expansion
are shown in Fig.~\ref{elagraph}.  One has two interaction channels,
which both proceed via the exchange of a scalar or pseudoscalar meson. The
typical cross section for these processes are of the order of millibarns.
The second class is represented by hadronization processes, $q\bar
q\otto\pi\pi$ \cite{su2hadron,su3hadron}, for which the Feynman diagrams
are shown in Fig.~\ref{hadgraph}. Here, one has an $s$-channel, which
proceeds via the exchange of a scalar resonance, and a $t$- and
$u$-channel.  Also here, the typical cross section is of the order
of millibarns.  For more details and plots of the cross sections, the
reader is referred to Refs.~\cite{elaste,su2hadron,su3hadron}.

A third class of processes, which cannot be treated by the scheme
detailled above, is the decay of pions into two quarks, for which
the Feynman diagram is shown in Fig.~\ref{decgraph}. This process is
possible if the pion mass fulfills the Mott condition $m_\pi>2m_q$
\cite{pion}. Here we proceed as follows: first we compute the mean
lifetime $\tau$ of the pion due to the process $\pi\to q\bar q$. Then
we decide with probability $1-\exp(-\Delta t/\tau)$, where $\Delta t$
is the time step of the calculation, if the particle decays. In this
case we again choose the direction of the outgoing particles randomly
in the meson rest frame and boost to the original frame.

In general, the quark and meson quasiparticle energies as well as
the scattering cross sections appearing in the collision integrals are
complicated functionals of the particle distributions, as can be seen from
Eqs.~(\ref{gap}), (\ref{pidis}) and the expressions for the cross sections
in Refs.~\cite{elaste,su2hadron,su3hadron}. An exact computation of these
quantities in a non-equilibrium situation is thus a tremendous numerical
task. Instead of making an exact computation at each time step, we thus
take a shortcut by defining an effective temperature. This is done as
follows: First, the gap equation (\ref{gap}) with the parametrization
(\ref{param}) inserted is solved exactly at each particle position
to give the quark mass field $m_q(\vec x, t)$. Then the effective
temperature is computed from the condition that the quark mass at the
respective point is equal to the equilibrium quark mass computed from
the effective temperature:
\be
m_q(\vec x, t) = m_q^{\rm eq}(T_{\rm eff}(\vec x, t)) \Punkt
\label{teffdef}
\ee
From this effective temperature we compute afterwards the pion mass
and the scattering cross sections, using the equilibrium expressions.
Note that $T_{\rm eff}$ is not the thermodynamic temperature, but rather
an auxiliary quantity for the computation of meson masses and cross
sections. It becomes, however, equal to the thermodynamic temperature
if thermal equilibrium is reached.

A similar procedure is used for the mass gradients, which appear in
Eq.~(\ref{transeq}). First, we compute $\vec\partial m_q$ from an
exact differentiation of Eqs.~(\ref{gap}), (\ref{param}). The pion mass
gradient is then computed from
\be \label{pigrad}
\vec\partial_x m_\pi(\vec x, t) = \frac{dm_\pi^{\rm eq}(T_{\rm eff})}{dT}
                 \left(\frac{dm_q^{\rm eq}(T_{\rm eff})}{dT}\right)^{-1}
                 \vec\partial_x m_q(\vec x, t) \Punkt
\ee
Note that the second factor on the right hand side of Eq.~(\ref{pigrad})
is negative, so that the effective force acting on pions due to the mean
field is opposite to the one acting on quarks.  With these prescriptions
at hand, Eqs.~(\ref{charact}) are solved to give the time evolution
of $\vec x_i(t)$, $\vec p_i(t)$ and thus the time evolution of the one
particle distributions.

The initial condition is chosen in the following way: In the beginning,
no pions are present, whereas the number of quarks and the number
of antiquarks are equal. The spatial center points of the Gaussians
representing the quarks $\vec x_i(0)$ are distributed homogeneously within
a sphere of radius $r_0$. The momenta $\vec p_i(0)$, on the other hand,
are distributed according to a Fermi distribution at a given temperature
$T_0$, which is a free parameter, and chemical potential $\mu = 0$.
The total number of quarks in the initial state is given by the
momentum integral over the Fermi distribution times the volume. Since
the NJL model is a low energy theory, the momentum distribution is cut
off at $p=\Lambda$.

\section{Numerical Results} \label{numsec}
In this section we show the numerical results of our simulation
program. All computations have been performed using the parameter
set $m_0=4$\,MeV, $G\Lambda^2=1.989$ and $\Lambda=820$\,MeV. The
width of the Gaussians in Eq.~(\ref{param}) was chosen to be $w=2$\,fm.

\subsection{Evolution of the Fireball}
To give an overview over the time evolution of the fireball, we first
describe one specific example of a system with initial radius $r_0=7$\,fm
and initial temperature $T_0=280$\,MeV. A snapshot of the expansion is
shown in Fig.~\ref{cartoon}. The upper part of Fig.~\ref{cartoon} shows
the initial state, which consists of a sphere filled up with quarks,
which are denoted by dark balls. At this time, the system contains
4684 quarks and antiquarks and has a kinetic energy density of
about 1.8\,GeV/fm$^3$. The bottom part of  Fig.~\ref{cartoon}
shows the same system 25\,fm$/c$ later. At this time, approximately
60\% of the quarks have been converted into pions, which are denoted by
the light balls. Since the NJL model is not confining, the remaining
40\% light quarks will not hadronize and remain present in the final
state.

Figure~\ref{massev} gives the time behaviour of the constituent quark
masses, averaged over all solid angles, as a function of $r$ at times
$t=0$, 5, 10, 15, 20 and 25\,fm$/c$.  At $t=0$, all quarks are sitting
within a sphere of radius 7\,fm. In the center of this sphere, the
particle density is maximal and the quark mass is thus low. Towards the
surface, the density drops due to the gaussian shape of the distribution
function and concomitantly the constituent quark mass rises. Note that
the quark mass is only known at the centre points $\vec x_i(t)$ of the
quark distribution functions and the plot ends thus at $r=7$\,fm. For
larger values of $r$, it would continuously rise towards the vacuum
value, thus forming a sort of `potential well', as was previously shown
in Ref.~\cite{trap}. At finite times, the system starts to expand due
to the thermal motion of the quarks. This expansion first depletes the
surface of the fireball, while the particle density in the centre stays
high, as can be seen from the plot at $t=5$\,fm$/c$ of Fig.~\ref{massev}.
At this time, some quarks have already gained their vacuum mass, while
the mass of those quarks, which are in the centre, is approximately
unchanged. At later times, when the depletion reaches the centre,
the `potential well' begins to flatten, until finally all quarks
have the vacuum mass. At this time, the interaction due to the mean
field ceases. This behaviour has been observed previously in
Refs.~\cite{trap,abada}.

The effective temperature, averaged over all solid angles, is shown in
Fig.~\ref{tempev}. Since this quantity is coupled to the constituent quark
mass via Eq.~(\ref{teffdef}), it shows qualitatively the same behaviour
as Fig.~\ref{massev}.  Initially, one has a high temperature region in
the centre of the fireball and zero temperature outside. As the expansion
progresses, the temperature drops in the centre. At $t=7$\,fm$/c$, the
temperature is lower than the Mott temperature at all space points.
At $t=25$\,fm$/c$, one arrives at a temperature of approximately 50\,MeV,
which is sufficiently close to zero to give no mean field contribution
any longer.

The result which can be drawn from this behaviour is that the high energy
density in the initial state leads to a large expansion velocity, which
in turn results in a short lifetime of the plasma. Although the increase
of the quark mass should make the expansion slower, this does not make a
large effect. Quantitatively, this can be seen from Fig.~\ref{schwerp},
where the mean radius is shown for three systems with identical initial
conditions, but different expansion mechanisms: one with full interaction,
one expanding according to a Vlasov equation (i.\,e. without collisions)
and one with no interaction at all. The curves for the interacting systems
lie below the one for the interaction free system, thus showing that the
growing quark mass leads to a slowing down of the expansion, which is,
however, not very strong. In the case of a fully interacting system,
some of the quarks are converted into pions, which are not slowed down,
so that in this case the mean radius is somewhat larger than in the
Vlasov case.  One may ask the question whether this short lifetime is
a generic result or only a consequence of the approximation of the NJL
approach. In view of the high initial energy density and the fact that
the quark masses are rather small, it is hard to see how an
expansion well below the speed of light can be achieved.

Figure~\ref{densev} shows the time behaviour of the angular averaged
particle densities, both for quarks and for pions. In the initial state
at $t=0$, only quarks are present. At later times, pions are produced and
thus contribute also to the particle density.

The time dependence of the particle multiplicities is shown in
Fig.~\ref{multifig}. Here it can be seen that the production of mesons
starts immediately after the beginning of the evolution. The pion
multiplicity rises steeply at $t=0$, while the quark multiplicity
goes down. This steep rise can be understood from the observation, that
the system is rather dense initially, thus giving rise to a high
collision rate. At later times, the multiplicity curves flatten.
There are two possible reasons for this flattening: either the
back reaction $\pi\pi\to q\bar q$ might become important due to the
large abundance of pions or the particle density becomes to low to
cause a further change of the multiplicities via collisions. To
answer this question, we show in Fig.~\ref{numcol} the number of
collisions per unit time, $dN_c/dt$, as a function of time. It can
be seen from this figure, that the vast majority of collisions
happen before $t=10$\,fm$/c$. If the reaction $\pi\pi\to q\bar q$
became important, the number of collisions would stay high, while
the multiplicities do not change any longer. Thus the reason for the
flattening of the multiplicity curves is the breakup of the fireball
rather than the approach to chemical equilibrium.

\subsection{Meson Production Mechanisms}
In the last subsection it has been pointed out, that approximately
half of the quarks in the system do not hadronize and remain present
in the final state. In order to investigate this behaviour further,
we have simulated systems with initial radii of 3, 4, 5, 6 and
7\,fm. For each initial radius, two runs have been performed. The
final multiplicities for each system, averaged over the two runs,
are given in Tab.~\ref{multitab}.  It can be seen from this table, that
the final state for all these runs contains about 50\% of quarks, but it
can also be seen, that there is a systematic trend towards a higher pion
abundance for larger systems. Whereas one has 42\% pions at $r_0=3$\,fm,
the pion abundance rises to 60\% at $r_0=7$\,fm. This can be interpreted
as an approach towards chemical equilibrium, which is nevertheless not
reached, since in an equilibrated cold system, only pions should be
present. The production of pions from quarks, however, takes a finite
time. The lifetime of a small expanding fireball is too short to hadronize
all quarks into pions without involving confinement. Nevertheless,
the lifetime rises with the initial radius, so that more quarks can
hadronize for a large system.

The hydrodynamical picture of the plasma expansion relies on the
assumption of local thermal equilibrium, which is the limit of infinitely
large cross sections of transport theory. In this case, the transition
from quark matter to hadronic matter takes place immediately after the
temperature drops below the critical temperature.  In an expanding finite
system, however, one has to take into account that the finite size of
the cross section leads to a finite mean free path and the hadronization
takes a finite time. To verify whether in our nonequilibrium, finite
particle number calculation a comparably strong relation between meson
production and temperature as in hydrodynamics is realized, we have
plotted in Fig.~\ref{thist} the creation probability of pions versus
the effective temperature. It can de seen that pions are predominantly
created at temperatures well {\em below\/} the Mott temperature within
a range of approximately 50\,MeV.  This figure should be compared to
Fig.~23 of Ref.~\cite{su3hadron}, where it has been shown that the mean
time for the conversion of quarks into mesons is minimal within the
same temperature range where Fig.~\ref{thist} shows a maximum of the
pion production.  This means that the mean free path is minimal below
the Mott temperature and quarks tend to hadronize here. We thus conclude
that the finite creation time of mesons cannot be neglected against the
expansion time scales of the plasma, i.\,e. effects of the finite mean
free path become important.

This behaviour is consistent with the density curves of
Fig.~\ref{densev}. Generally one observes, that in the early stages,
when the system is still hot in the center, the meson density is maximal
at a finite value of $r$. At later times, when the temperature in the
centre drops below the Mott temperature, one has the situation, that
the meson density follows the quark density.

\subsection{Density Fluctuations}
The initial conditions used above are, besides statistical fluctuations
in the distribution of the particles, spherically symmetric. We see,
however, in Fig.~\ref{dflukt}, where we plot the summed particle densities
for quarks and pions along the coordinate axes, both for the initial and
final state, that these small fluctuations become amplified during the
expansion and that the system does not at all keep the symmetry. This is
understandable because the increase of the parton mass with decreasing
density is equivalent to an attractive force between them. Therefore
partons tend to cluster and the initial fluctuations are the seeds for
this clusterization. This observation challenges, like the finite mean
free path effects, the possibility to employ hydrodynamical approaches,
which are only applicable if initial fluctuations are damped.

\subsection{Variation of Initial Conditions}
\subsubsection{Cylindrical Initial Conditions} \label{cynsec}
The initial conditions given in Sec.~\ref{modelsec} are not the ones
one would expect in a heavy ion collision. To come closer to the
experimental situation, we made one run with
cylindrical initial conditions. The spatial center points in this run are
equally distributed within a cylinder which, in longitudinal direction,
extends from $-z_{\rm max} \le z \le z_{\rm max}$ with $z_{\rm max}=2$\,fm
and in transverse direction has a radius of 6\,fm. The distribution of the
momenta perpendicular to the cylinder axis was chosen to be gaussian
\be
p(p_x, p_y) \sim \exp\left(-\frac{p_x^2 + p_y^2}{2(\delta p)^2} \right)
\ee
with width $\delta p=250$\,MeV. The distribution along the cylinder
axis, however, was chosen to depend linearly on $z$ in such a fashion,
that $p_z=\pm\Lambda$ is reached at the cylinder faces at $z=\pm z_{\rm
max}$:
\be
p(p_z) = \delta(p_z - \Lambda z / z_{\rm max})
\Punkt
\ee
The total number of particles was 2000, which gives approximately
the same particle density in the initial state as for the other runs.
After 30\,fm$/c$, 1126 quarks have been converted into mesons, which
corresponds to 56\% of the total particle number. In Fig.~\ref{multi2},
we show the time dependence of the multiplicity. This figure shows
the same trend as Fig.~\ref{multifig}. The geometry
does thus not have a large impact on the particle multiplicities.
Figure \ref{rapper} shows the distribution of the center points $\vec
p_i(t)$ in rapidity both for the initial and the final state. At $t=0$,
one sees a distribution with two maxima at $y=\pm 1.1$. At $t=30$\,fm$/c$,
the quark rapidity distribution has become flatter and narrower, while
the meson distribution is flat at midrapidity, but has a larger width than the
quark distribution. This comes from the fact that quarks are slowed
down during to the expansion and thus move to lower rapidities. Mesons,
on the other hand, have a lower mass than quarks at the time of their
creation and thus obtain a higher momentum as the incoming quarks. This
leads to a spread of their rapidity distribution.

Figure \ref{pty} shows the mean transverse momentum as a function of
rapidity. Here one encounters the same behaviour as for the particle
density, in that the distribution for the mesons is wider than for
the quarks. The geometry of the expansion does thus not influence the
$p_T$ spectra, but shifts the quarks to midrapidity. It is not very
astonishing that mesons have the same mean transverse momentum in
midrapidity, because the process $q\bar q\to\pi\pi$ leads to about the
same value of $p_T$.

\subsubsection{Mixed Temperature Initial Condition}
In view of the short lifetime of the fireball, the question arises
whether the lifetime is long enough in order to establish thermal
equilibrium. For this reason we have studied a spherically symmetric
system, in which half of the particles with $r>r_0/2$ obtained a momentum
according to a temperature of 180\,MeV (`cold' particles), whereas the
other particle momenta were initialized with a temperature of 280\,MeV
(`hot' particles). If thermal equilibrium was established, the `cold'
particles would obtain the same energy as the `hot' particles in the
course of the expansion.  The result is shown in Fig.~\ref{equi}, where
we plot the energy per particle for both subsystems as a function of
time. It can be seen, that the energy of the `cold' particles approaches
that of the `hot' particles, but stays below. One has thus no complete
thermal equilibration.  This reflects the already discussed observation
that the expansion velocity is very large, too large to produce a global
equilibrium.

\subsection{Variation of Hadronization Cross Sections}
Since the NJL model is nonconfining, one is faced with the question,
if a sufficiently enlarged hadronization cross section could lead to a
system, which complete hadronizes into mesons. We thus took the initial
conditions of Sec.~\ref{cynsec} and made several calculations, in which we
increased all cross sections of the hadronization type, i.\,e. the ones
for $q\bar q\to\pi\pi$ as well as those for the back reaction $\pi\pi\to
q\bar q$, artificially by a constant factor. The result is given in
Table~\ref{hadrotab}, where we give the particle abundancies in the
final state for these runs.  One sees that the increased hadronization
cross section leads to an increased pion abundance in the final state,
but nevertheless not to a complete hadronization.

\section{Summary and Conclusions} \label{sumsec}
In this paper, we have studied the expansion and hadronization of systems,
which interact according to a chirally symmetric Lagrangian. These
calculations cannot serve to reproduce experimental data, since the
underlying interaction is not confining. Since, however, chiral symmetry
is an decisive feature of hadron phenomenology, it serves nevertheless to
address qualitatively several questions, which we believe to persist
to more realistic scenarios. One of these points concerns the question
of collision and expansion time scales. We find that the expansion and
cooling progresses relatively fast, so that the finite time, which is
needed to produce particles, cannot be neglected. We obtain the result,
that chemical and even thermal equilibrium are not necessarily established
during the expansion.

Another interesting point is the production mechanism of mesons, which
is studied here for the first time. We find that mesons are created
mainly within a temperature range of approximately 50\,MeV below the
Mott temperature. This effect can be traced back to a minimum of the
mean free path. This feature might persist to a confining scenario,
because it is hard to see how, even in a confining theory, pions could
be created instantaneously.  We also find that mesons are created in the
bulk of the fireball rather than at its surface and that at late times
the meson density follows the quark density.

Starting from an initially spherically symmetric configuration, we observe
that this configuration does not necessarily stay symmetric, but rather
tends to form local density maxima at late times. All these last three
statements challenge the applicability of hydrodynamical models.

Using a cylindrically symmetric initial configuration, which might be
more realistic to encounter in an experimental situation, we find a
flat rapidity distribution for the produced particles. The rapidity
dependence of the transverse momentum is also flat. We observe, that
in the final state mesons cover a wider range in rapidity than quarks,
which is due to the quark mean field.

More interesting than the $SU(2)$ case, which is presented here, is
the extension to $SU(3)$. There it will be possible to verify the
theoretical concepts for the production of strangelets or strangeness
destillation. This will be the subject of an upcoming publication.

\begin{table}
\begin{tabular}{c||c|c|c|c|c||c|c|c|c}
$r_0$ (fm) & $N_q$ & $N_u$ & $N_{\bar u}$ & $N_d$ & $N_{\bar d}$ & $N_{\pi}$
& $N_{\pi^0}$ & $N_{\pi^+}$ & $N_{\pi^-}$ \\
\hline
3 & 216 & 51.5 & 56.5 & 52 & 56 & 157 & 52.5 & 49.5 & 55 \\
4 & 450 & 106.5 & 118.5 & 107.5 & 117.5 & 422 & 155 & 133 & 134 \\
5 & 821 & 203.5 & 207 & 200 & 210.5 & 887.5 & 310 & 290.5 & 287 \\
6 & 1283 & 320.5 & 321 & 319 & 322.5 & 1688.5 & 578 & 556 & 554.5 \\
7 & 1931 & 480 & 485.5 & 482.5 & 483 & 2883.5 & 989 & 946 & 948.5 \\
\end{tabular}
\caption[]{Multiplicity distribution in the final state as a function
           of the system size. The columns $N_q$ and $N_\pi$ give the sum
           over all quarks and all pions, respectively.}
\label{multitab}
\end{table}

\begin{table}
\begin{tabular}{c||c|c|c}
$\sigma_h/\sigma_h^0$ & $n_q$ & $n_\pi$ & $n_\pi/(n_q + n_\pi)$ \\
\hline
1  & 874 & 1126 & 0.563 \\
2  & 708 & 1292 & 0.646 \\
5  & 538 & 1462 & 0.731 \\
10 & 420 & 1580 & 0.790 \\
\end{tabular}
\caption[]{Particle multiplicity and ratio of pions to all particles
           for several calculations with artificially increased
           hadronization cross sections.}
\label{hadrotab}
\end{table}

\begin{figure}
\caption[]{Feynman graph for the irreducible pseudoscalar polarization.
           Solid lines denote constituent quarks.}
\label{polarfig}
\end{figure}

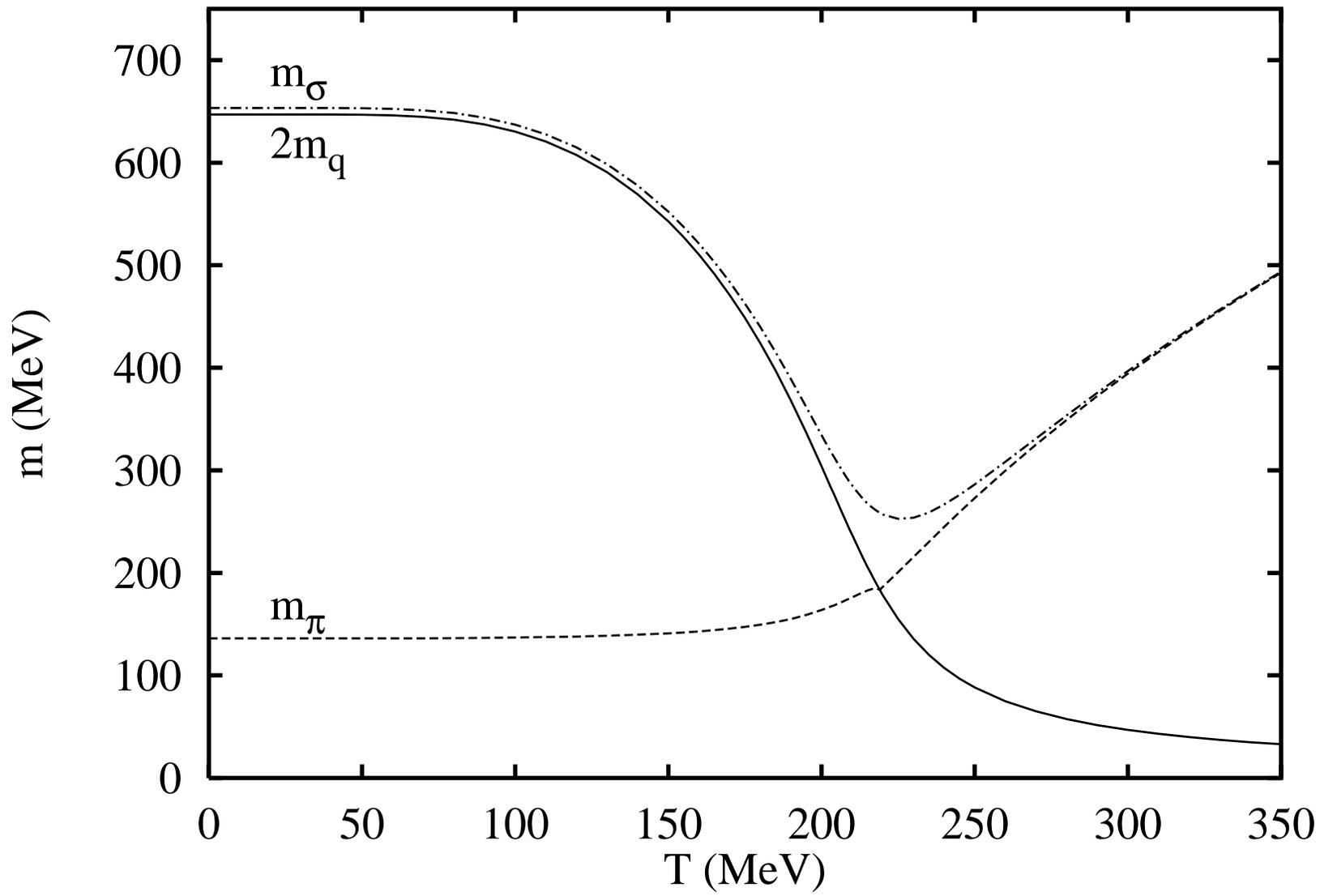
\begin{figure}
\caption[]{Mass spectrum of the $SU(2)$ NJL model in thermal equilibrium
           as a function of temperature. Solid line: quark mass times two,
           dashed line: pion mass, dot-dashed line: sigma mass.}
\label{massfig}
\end{figure}

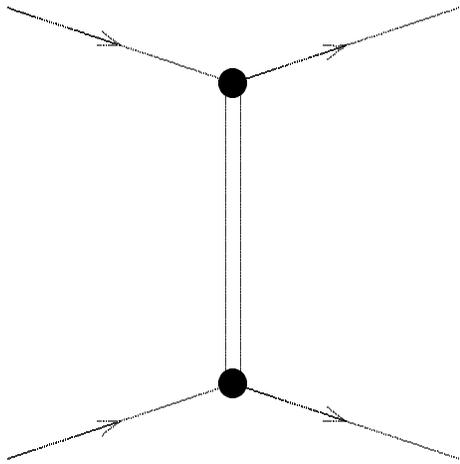
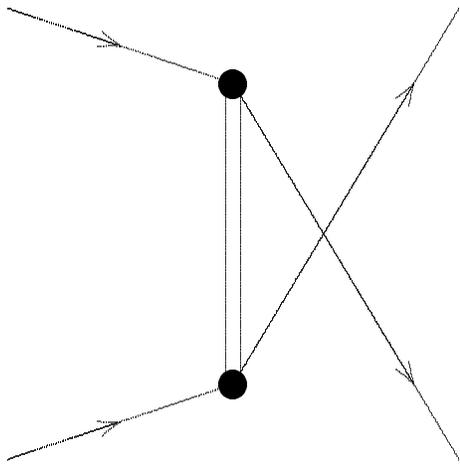
\begin{figure}
\caption[]{Feynman graphs for the process $qq\to qq$. Single lines
           denote quarks, double lines mesons. The graphs for the process
           $q\bar q\to q\bar q$ are obtained from these by a rotation
           around 90 degrees.}
\label{elagraph}
\end{figure}

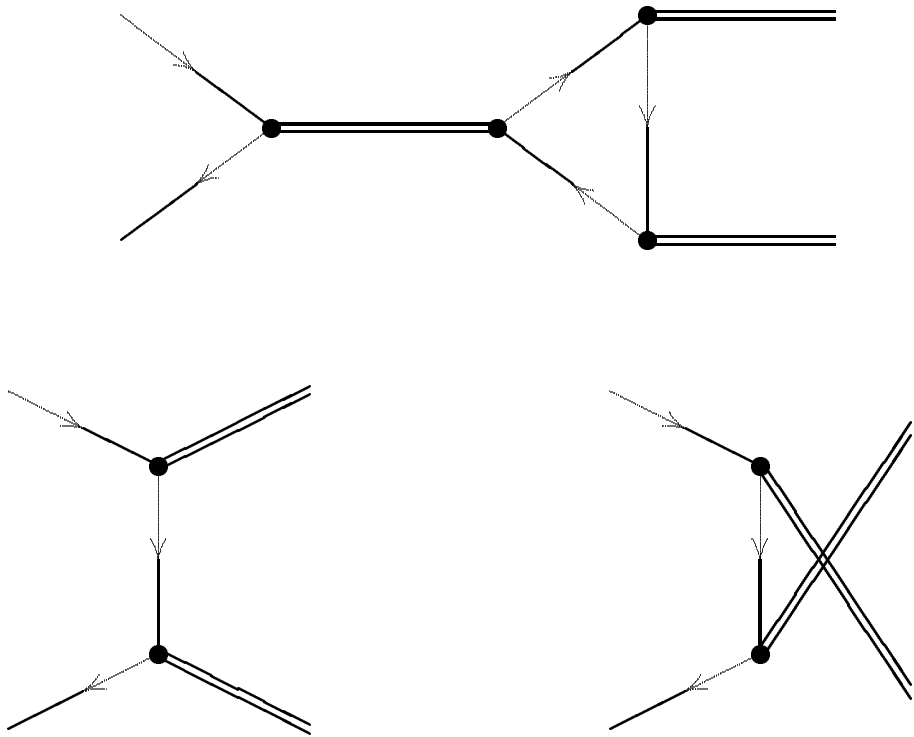
\begin{figure}
\caption[]{Feynman graphs for the process $qq\to\pi\pi$. Single lines
           denote quarks, double lines mesons.}
\label{hadgraph}
\end{figure}

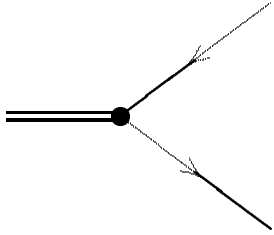
\begin{figure}
\caption[]{Feynman graphs for the process $\pi\to q\bar q$. Single lines
           denote quarks, double lines mesons.}
\label{decgraph}
\end{figure}

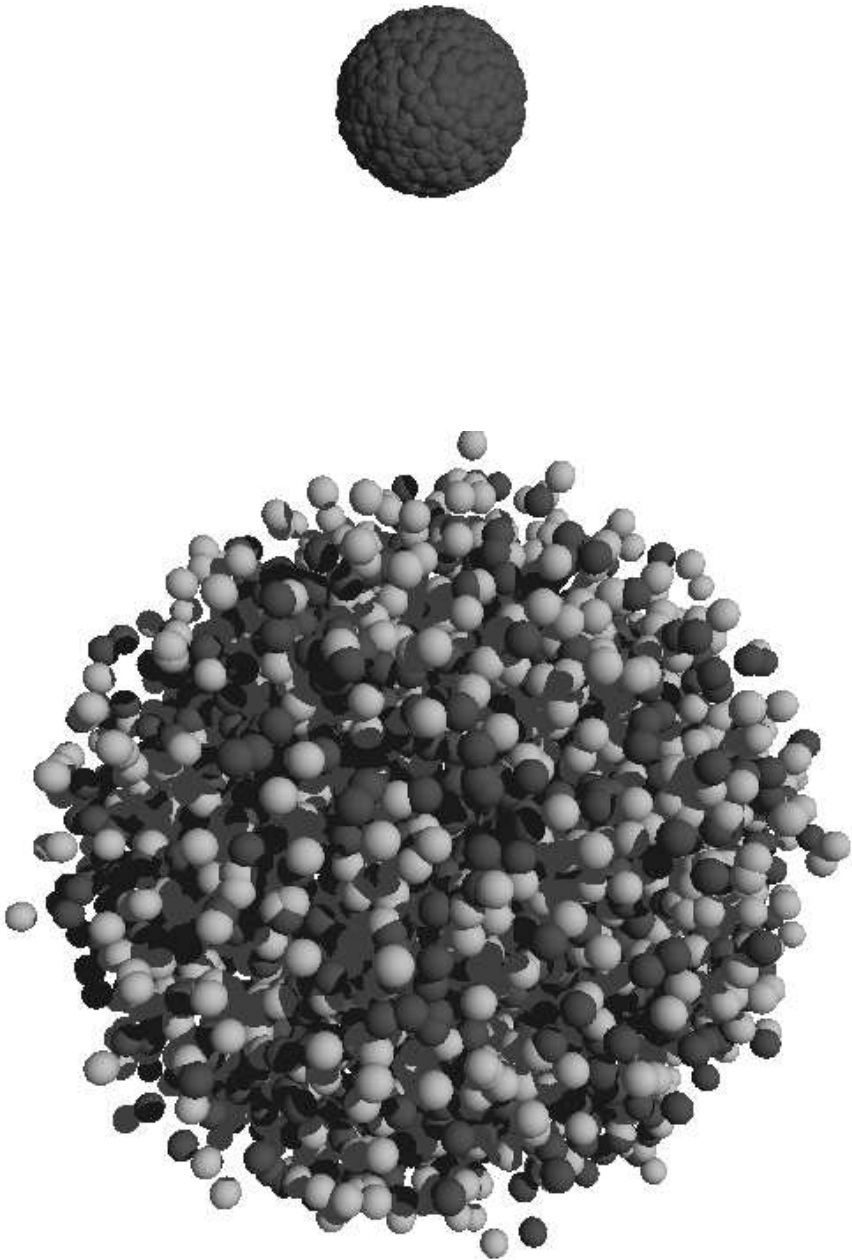
\begin{figure}
\caption[]{Snapshot of the expansion of the fireball. Top: the system
           at $t=0$, bottom: the same system at $t=25$\,fm$/c$. Dark
           balls denote quarks, light balls denote pions.}
\label{cartoon}
\end{figure}

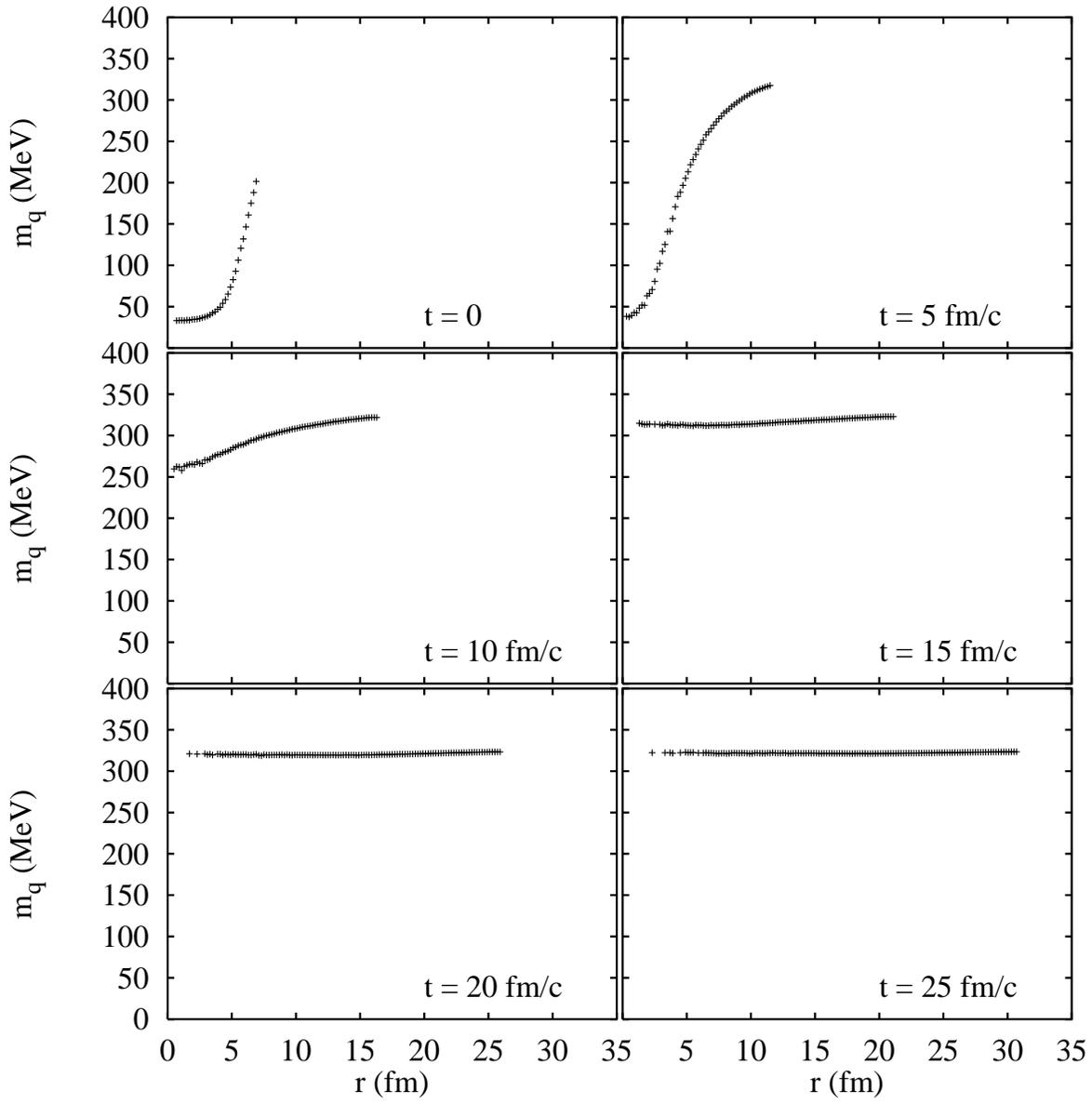
\begin{figure}
\caption[]{The constituent quark mass, averaged over all solid angles,
           as a function of $r$ at different times.}
\label{massev}
\end{figure}

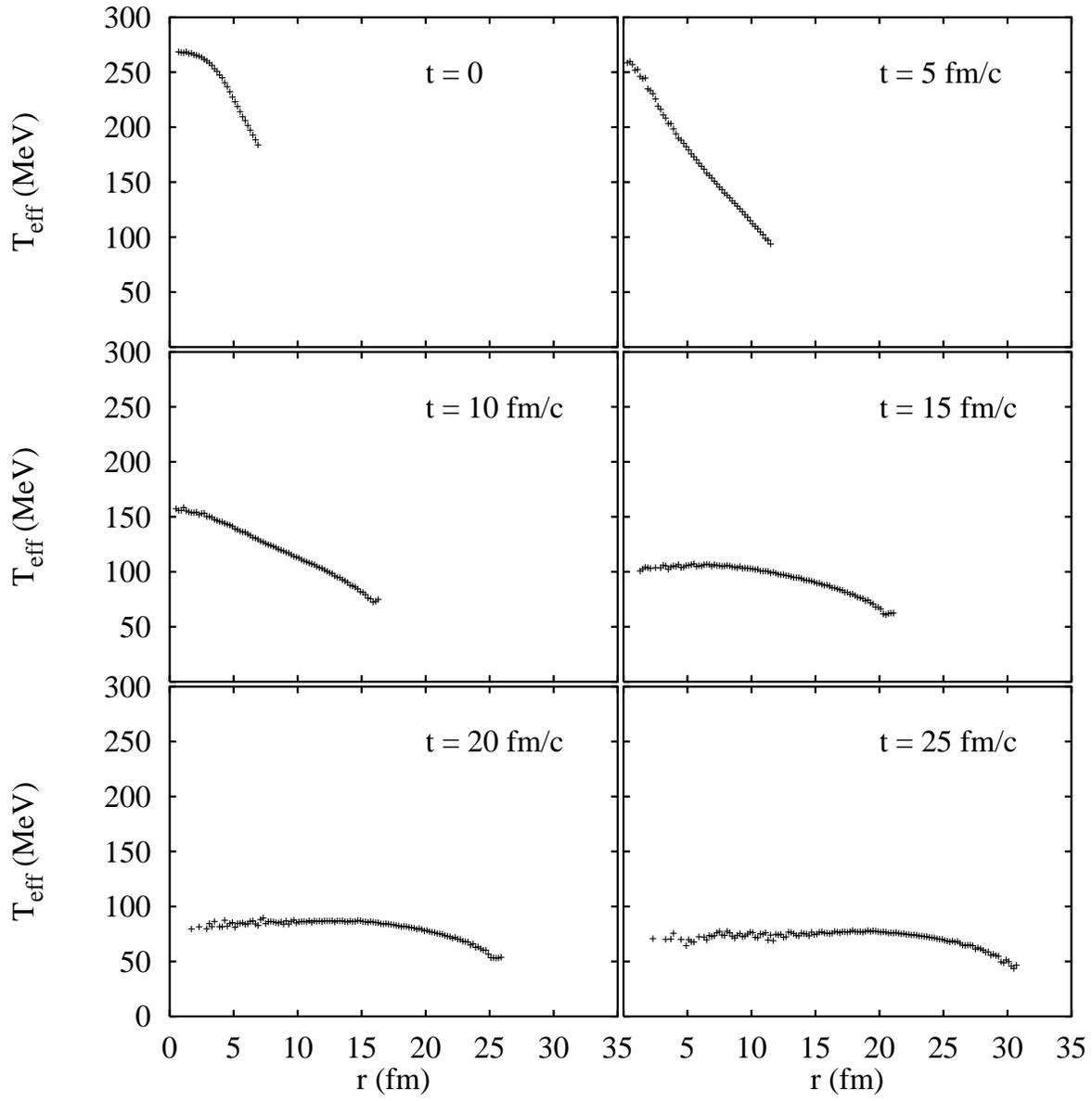
\begin{figure}
\caption[]{The effective temperature, averaged over all solid angles,
           as a function of $r$ at different times.}
\label{tempev}
\end{figure}

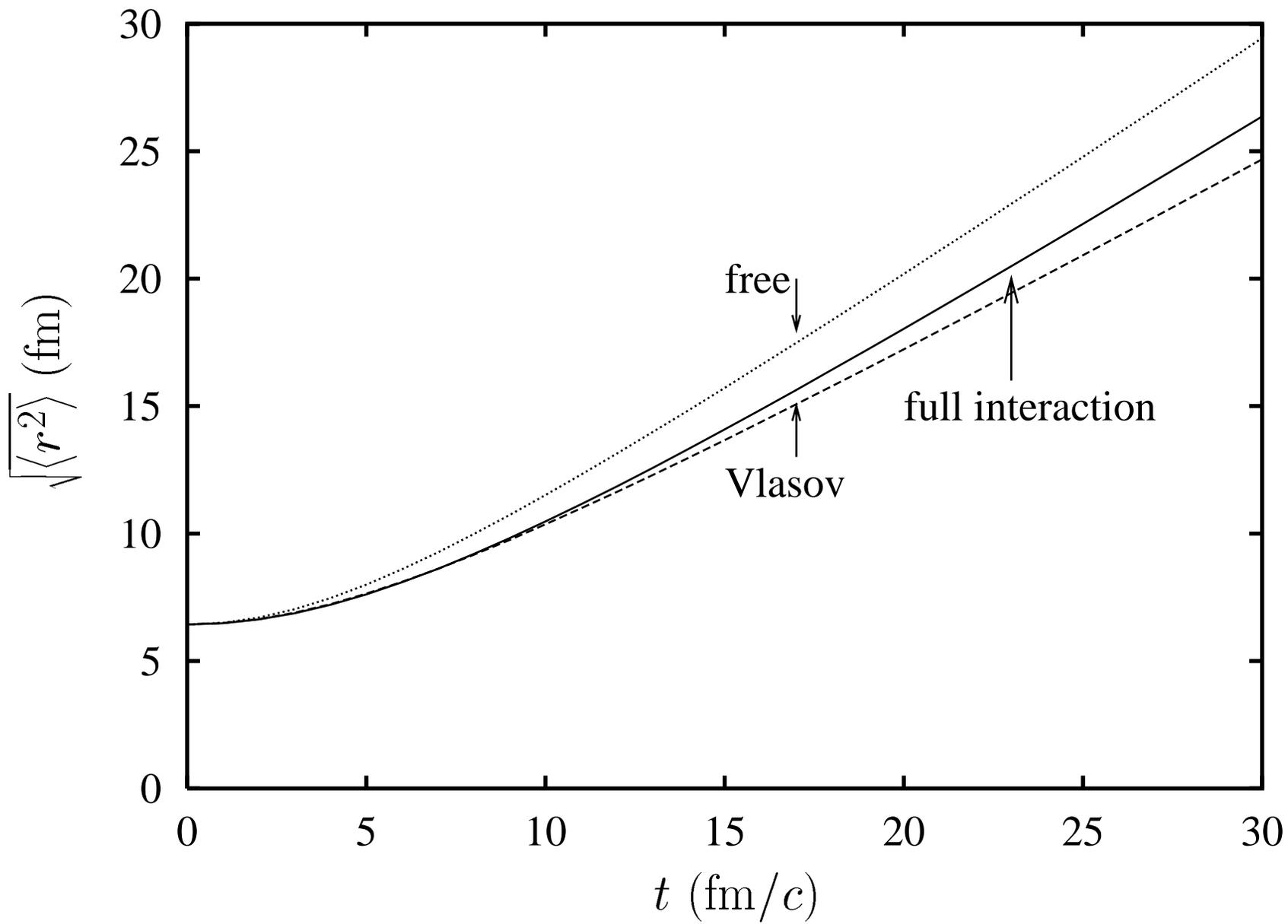
\begin{figure}
\caption[]{Mean radius of the fireball as a function of time for different
           expansion scenarios. Full line: full interaction, dashed line:
           Vlasov expansion, dotted line: free expansion.}
\label{schwerp}
\end{figure}

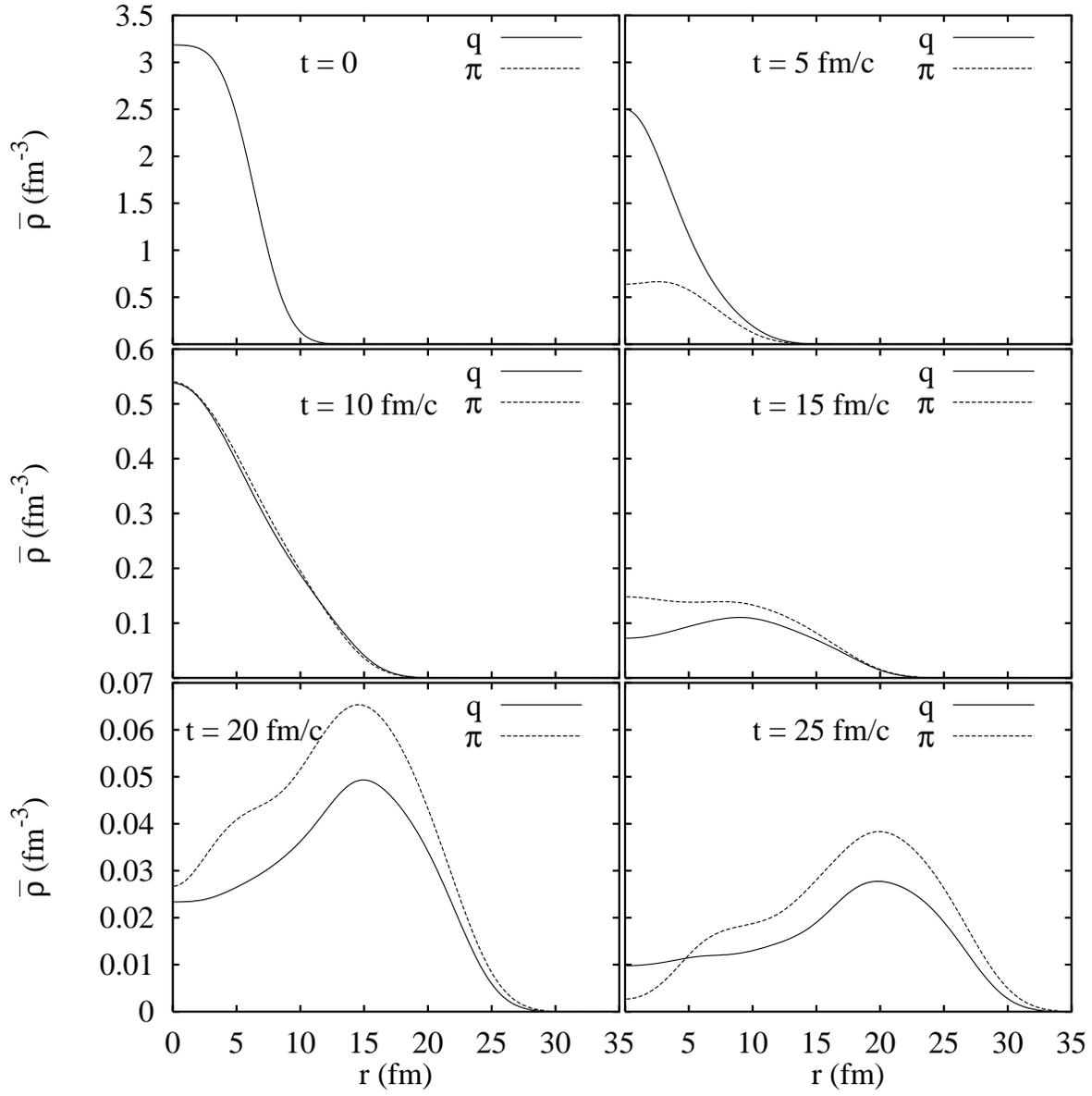
\begin{figure}
\caption[]{The particle density, averaged over all solid angles,
           as a function of $r$ at different times. Solid lines: quarks,
           dashed lines: pions. Note the different scales on the vertical
           axes of the three rows.}
\label{densev}
\end{figure}

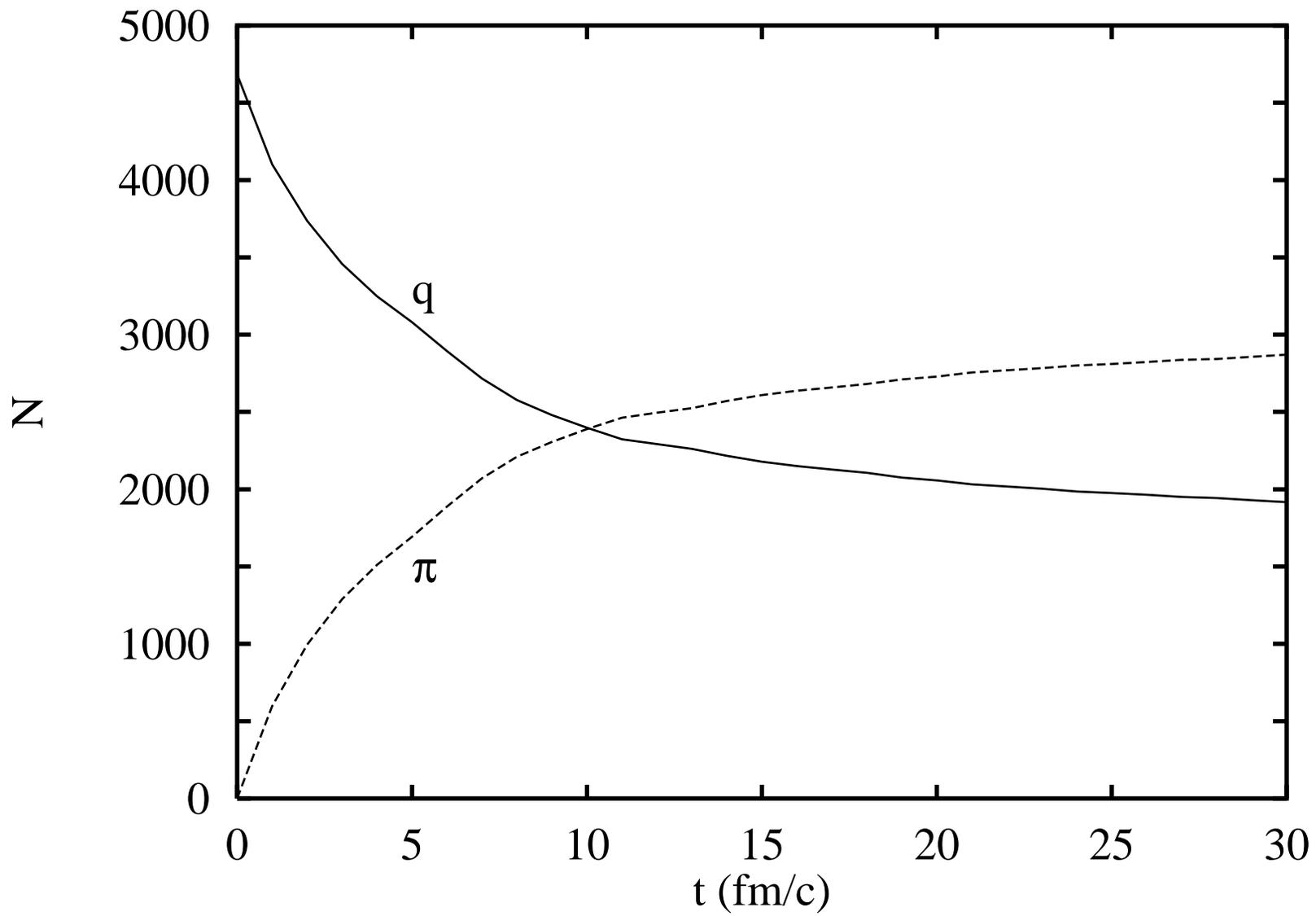
\begin{figure}
\caption[]{The particle multiplicity as a function of time. Solid line:
           quarks, dashed line: pions.}
\label{multifig}
\end{figure}

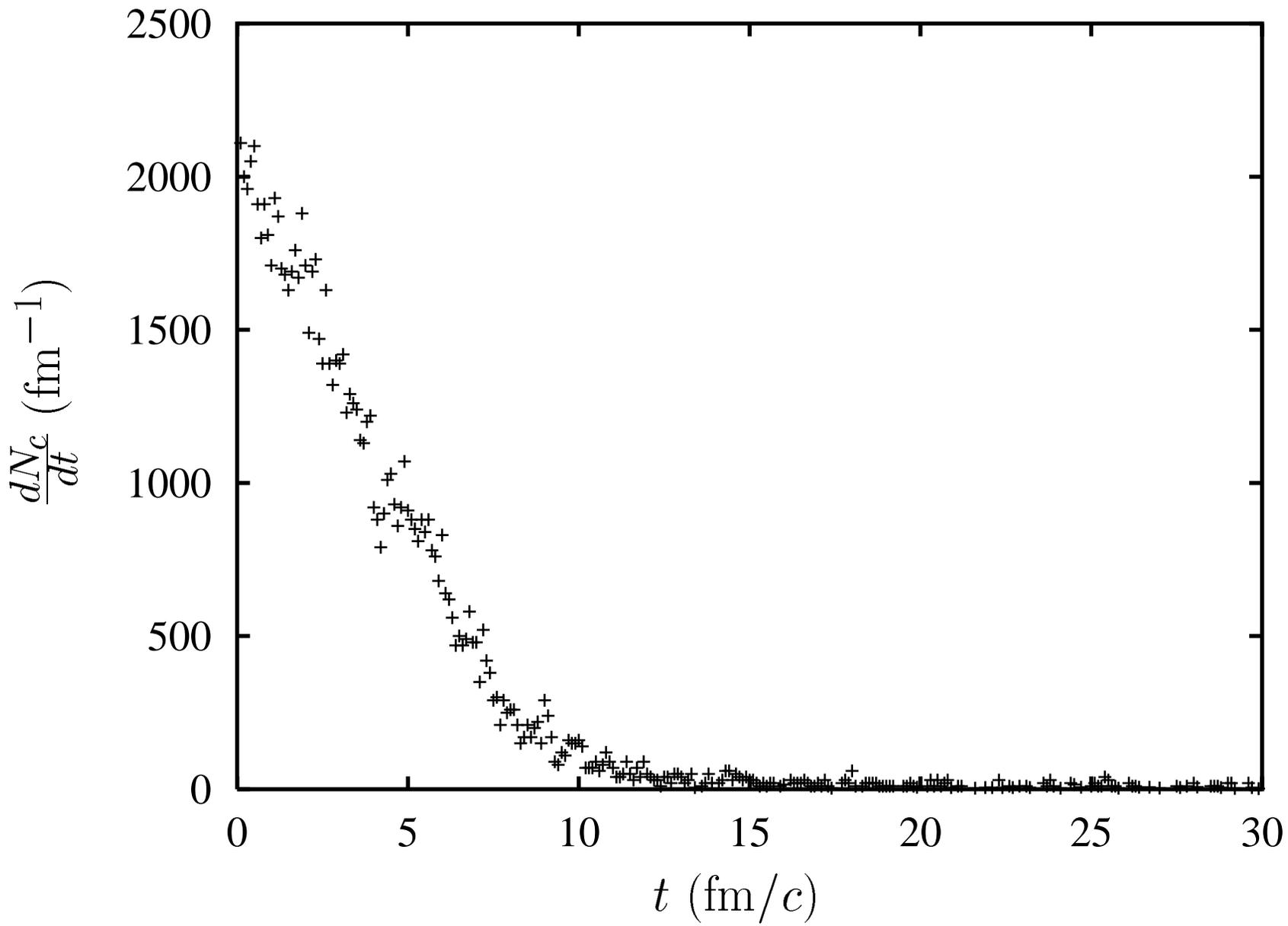
\begin{figure}
\caption[]{Time dependence of the total number of collisions per time
           interval.}
\label{numcol}
\end{figure}

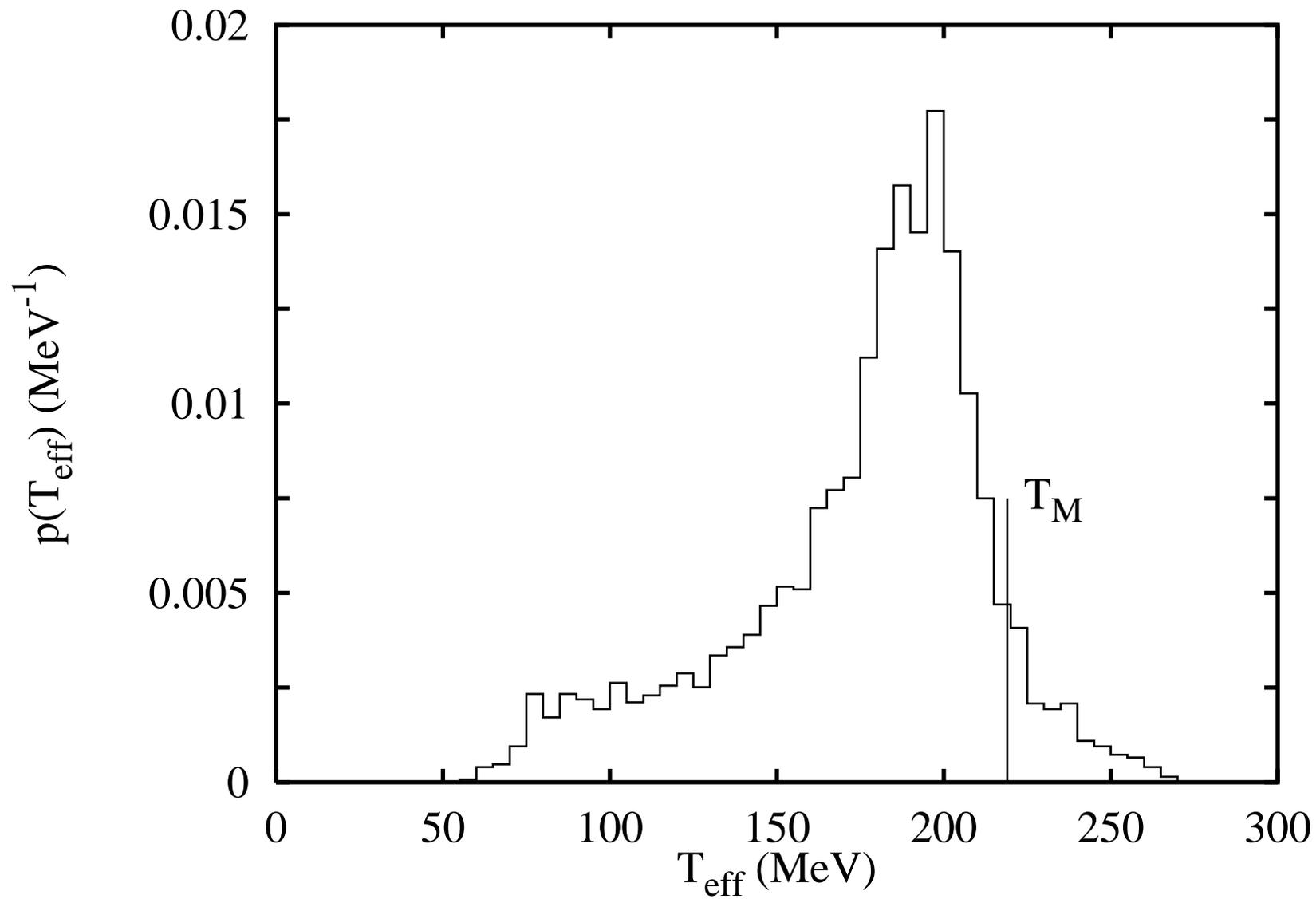
\begin{figure}
\caption[]{Creation probability of pions as a function of the effective
           temperature. The Mott temperature is marked by the vertical
           line.}
\label{thist}
\end{figure}

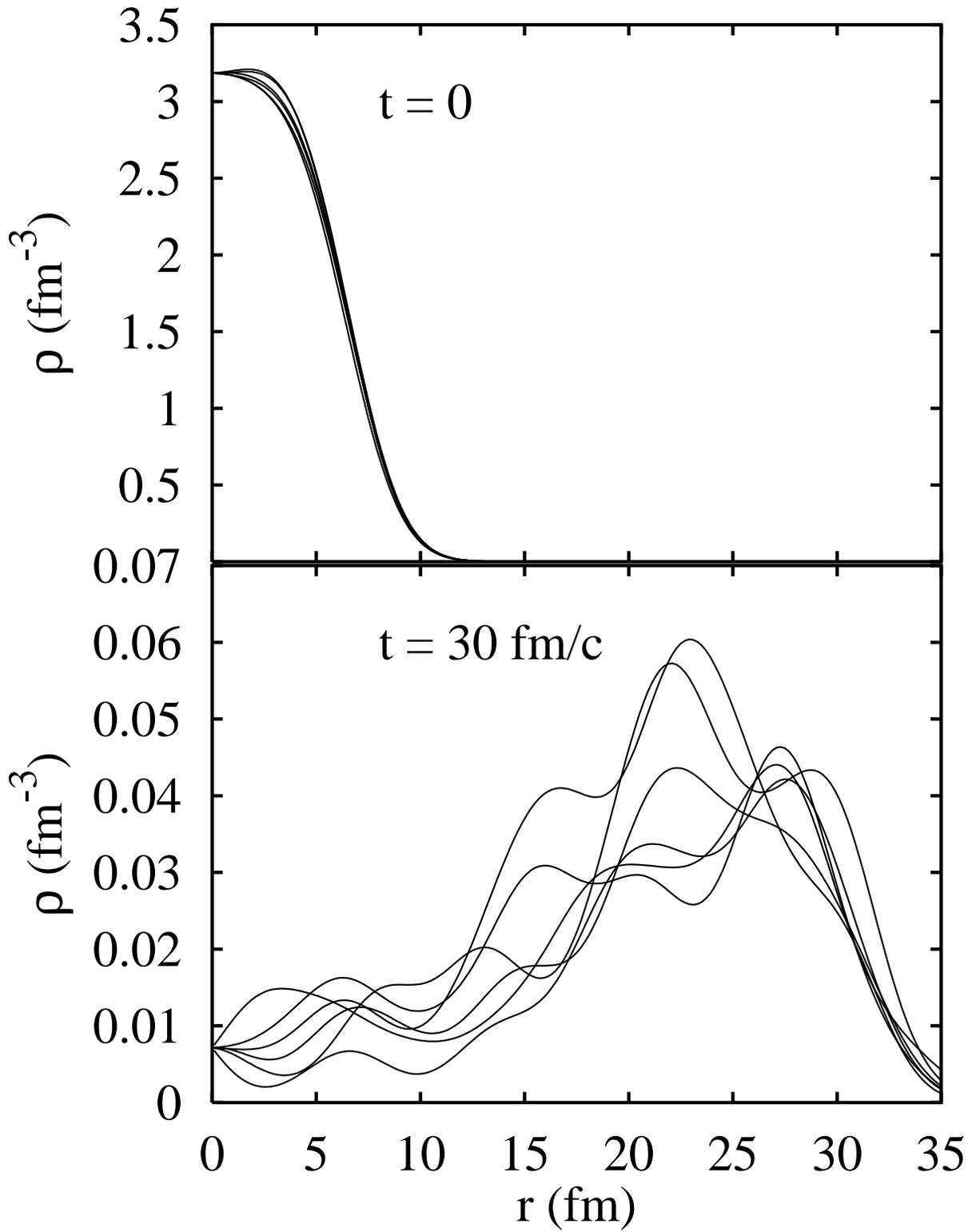
\begin{figure}
\caption[]{Particle density summed over quarks and mesons along the
           coordinate axes. Upper panel: $t = 0$, lower panel:
           $t=30$\,fm$/c$.}
\label{dflukt}
\end{figure}

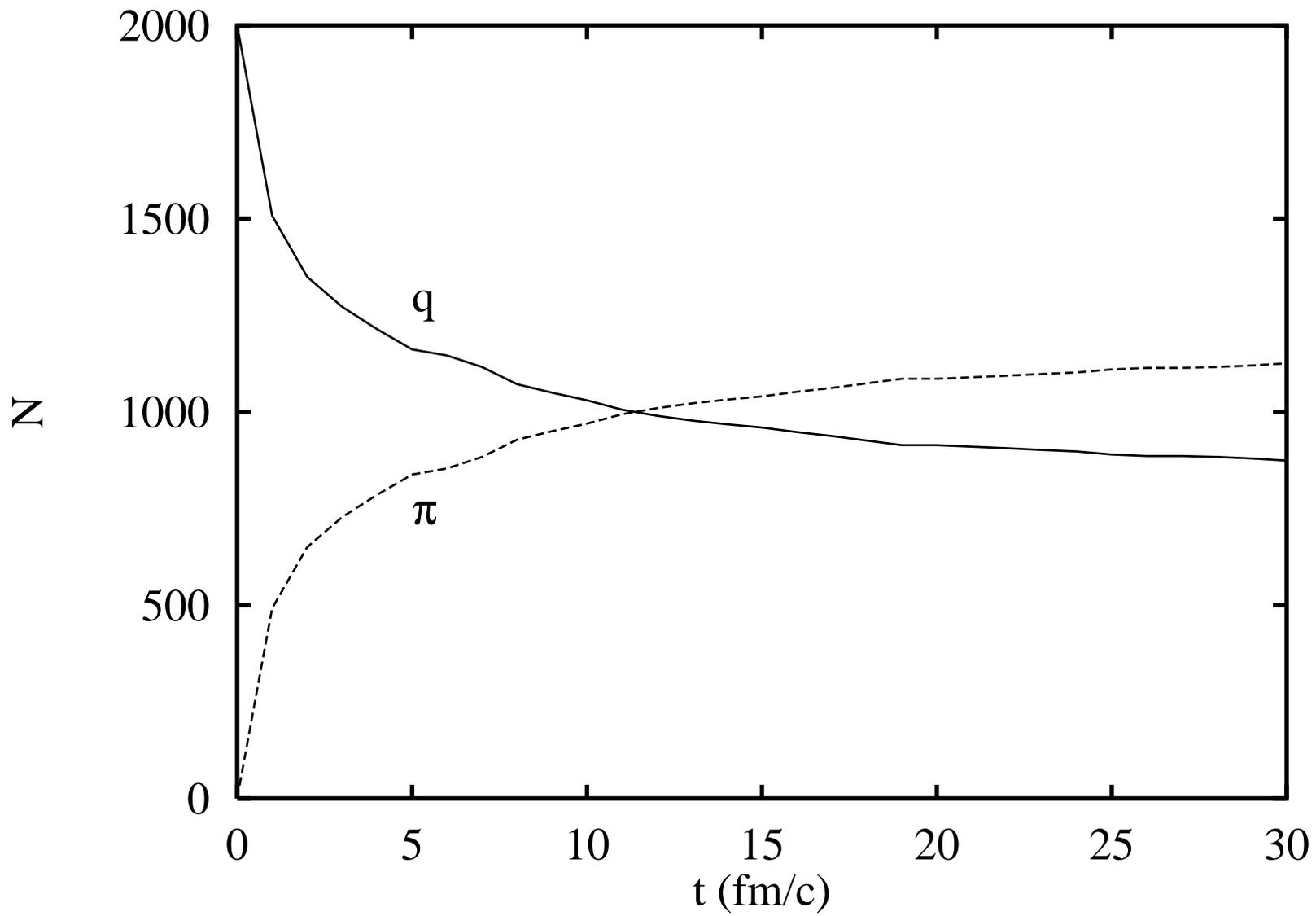
\begin{figure}
\caption[]{Time dependence of the multiplicity for a system with cylindrically
           symmetric initial conditions. The solid line denotes quarks, the
           dashed line pions.}
\label{multi2}
\end{figure}

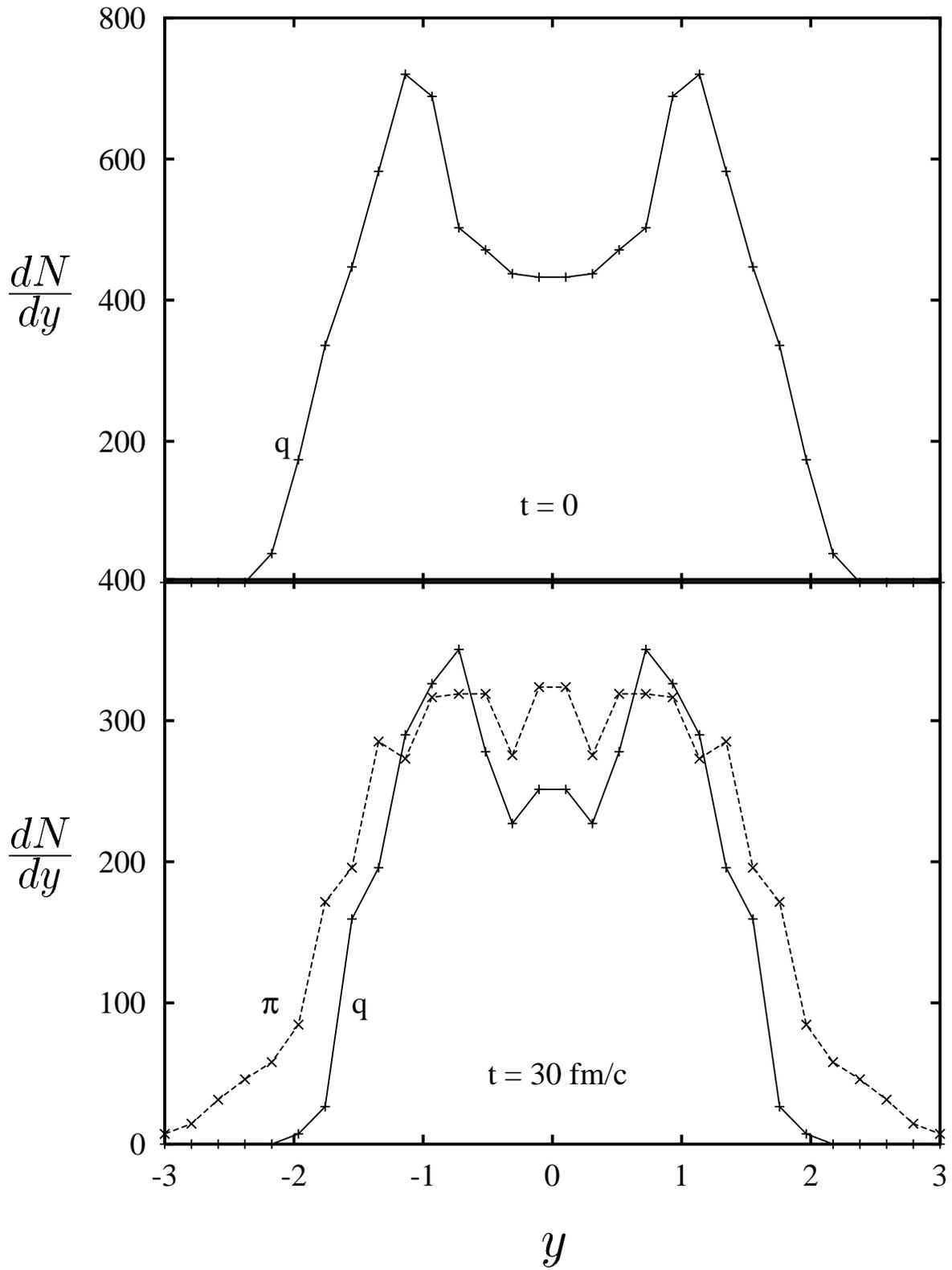
\begin{figure}
\caption[]{Distribution of center points in rapidity for a system with
           cylindrically symmetric initial conditions. Upper panel:
           quark distribution in the initial state, lower panel: quark
           ($+$) and pion ($\times$) distribution in the final state. The
           lines are a guide to the eye only.}
\label{rapper}
\end{figure}

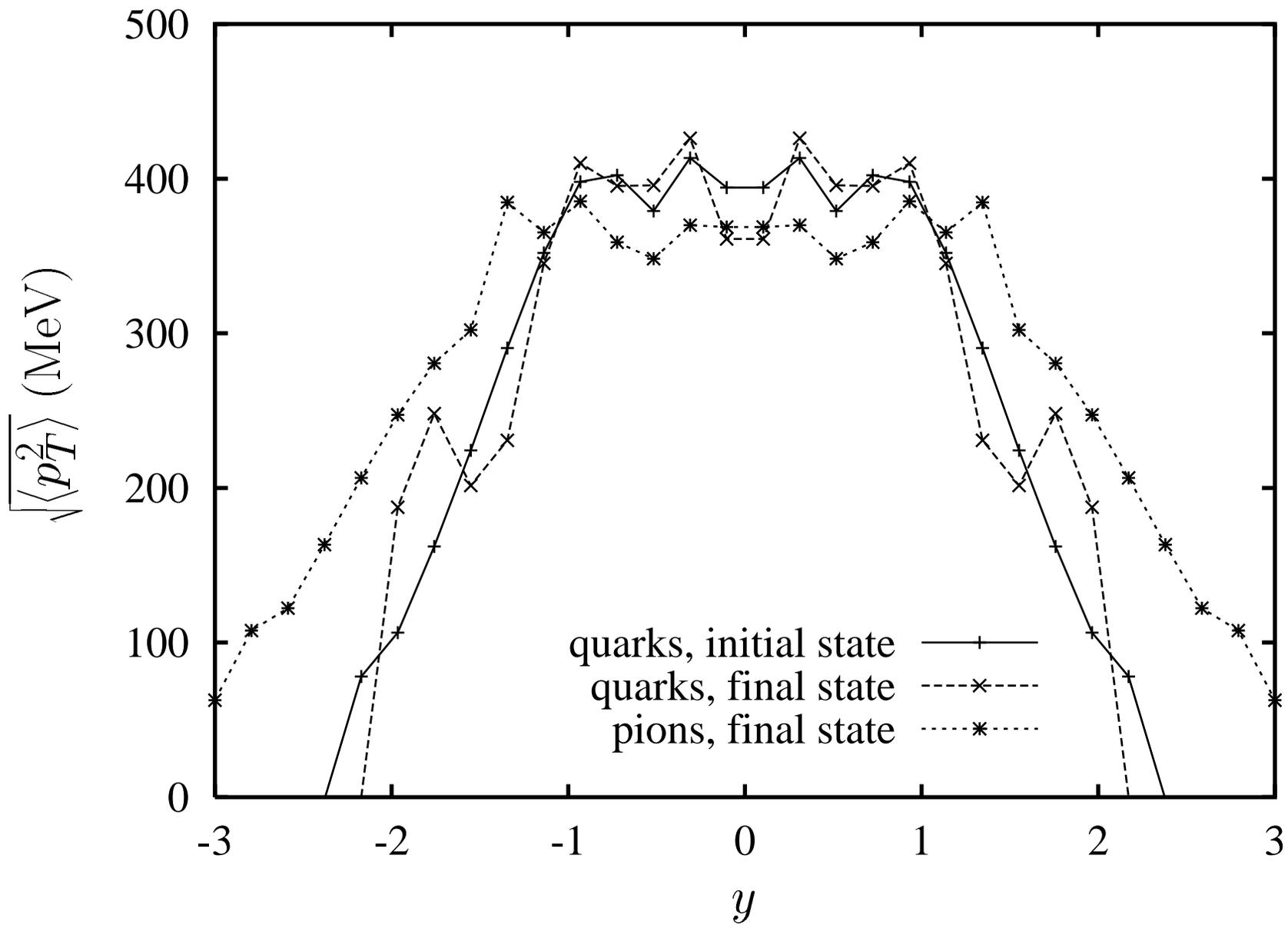
\begin{figure}
\caption[]{Rapidity dependence of the mean transverse momentum for quarks
           in the initial state ($+$), quarks in the final state ($\times$)
           and pions in the final state ($*$).}
\label{pty}
\end{figure}

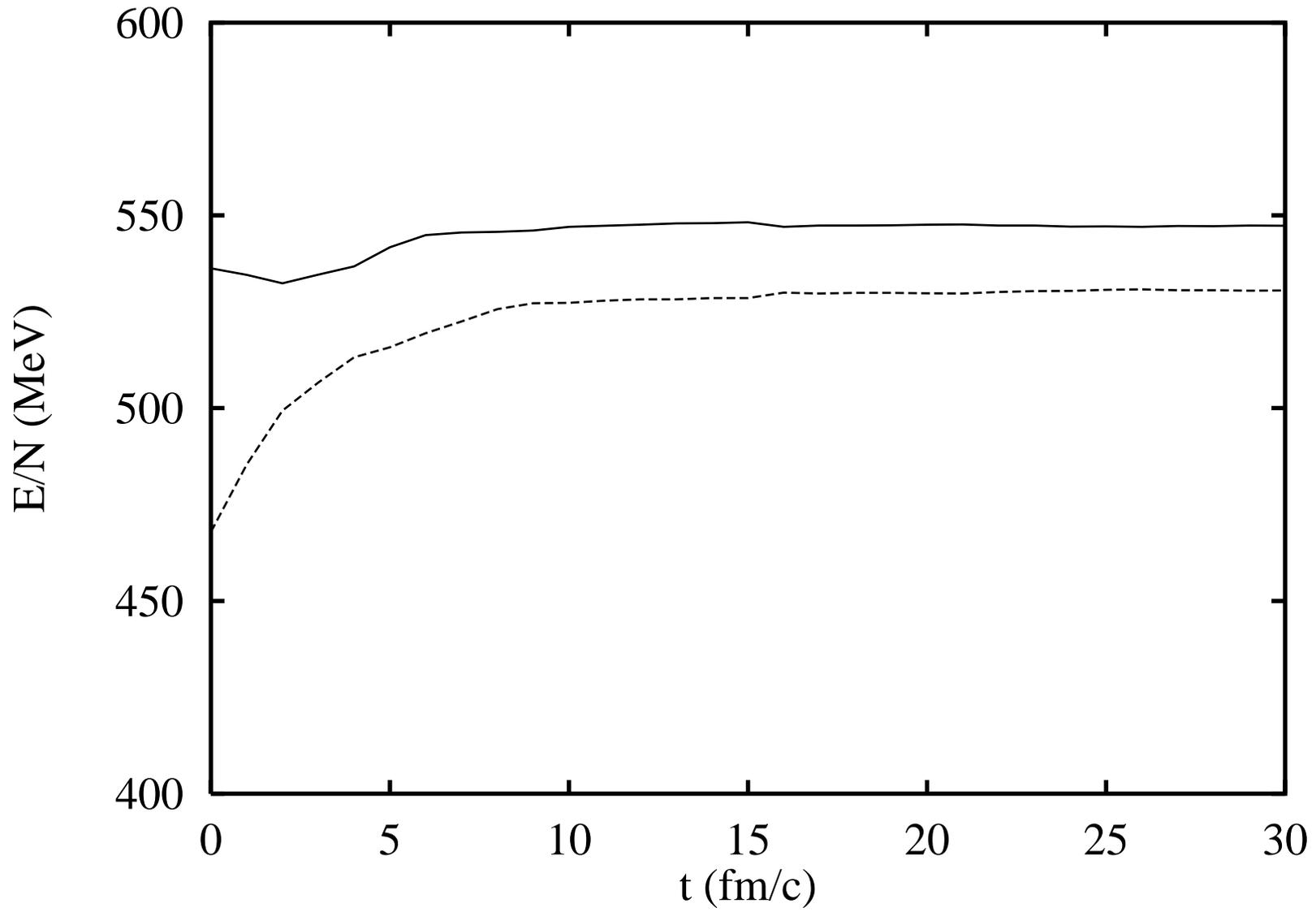
\begin{figure}
\caption[]{Energy density per particle for a fireball, in which a part of
           the quark momenta has been initialized according to a lower
           temperature. Solid line: `hot' particles, dashed line: `cold'
           particles. Please note the broken scale on the vertical axis.}
\label{equi}
\end{figure}


\begin{references}
\bibitem{qm96}
        {\it Proceedings of Quark Matter '96\/}, Eds.
        P.~Braun--Munzinger, H.\,J.~Specht, R.~Stock and H.~St{\"o}cker,
        Nucl. Phys. A {\bf 610}, 1c (1997);
        {\it Proceedings of Quark Matter '97\/}, Nucl. Phys. A, in press.
\bibitem{hydro}
        J.\,D.~Bjorken, Phys. Rev. D {\bf 27}, 140 (1983);
        J.~Bolz, U.~Ornik, M.~Pl\"umer, B.\,R.~Schlei and R.\,M.~Weiner,
        Phys. Rev. D {\bf 47}, 3860 (1993);
        D.\,H.~Rischke and M.~Gyulassy, Nucl. Phys. A {\bf 597}, 701 (1996).
\bibitem{nambu}
        Y.~Nambu and G.~Jona-Lasinio, Phys. Rev. {\bf 122}, 345 (1961);
        Y.~Nambu and G.~Jona-Lasinio, Phys. Rev. {\bf 124}, 246 (1961).
\bibitem{sandi}
        S.\,P.~Klevansky, Rev. Mod. Phys. {\bf 64}, 649 (1992);
        T.~Hatsuda and T.~Kunihiro, Phys. Rep. {\bf 247}, 221 (1994).
\bibitem{zhawi}
        W.-M.~Zhang and L.~Wilets, Phys. Rev. C {\bf 45}, 1900 (1992).
\bibitem{wojtek}
        W.~Florkowski, J.~H\"ufner, S.\,P.~Klevansky and L.~Neise,
        Ann. Phys. (NY) {\bf 245}, 445 (1995).
\bibitem{ogu}
        S.\,P.~Klevansky, A.~Ogura and J.~H\"ufner, Ann. Phys. (NY)
        {\bf 261}, 37 (1997).
\bibitem{trap}
        P.~Rehberg and J.~H\"ufner, Nucl. Phys. A {\bf 635}, 511 (1998).
\bibitem{pion}
        P.~Rehberg, Phys. Rev. C {\bf 57}, 3299 (1998).
\bibitem{abada}
        A.~Abada and J.~Aichelin, Phys. Rev. Lett. {\bf 74}, 3130 (1995);
\bibitem{bot}
        L.~Bot and J.~Aichelin, J. Phys. G {\bf 23}, 1947 (1997).
\bibitem{chpt}
        J.~Gasser and H.~Leutwyler, Ann. Phys. (NY) {\bf 158}, 142 (1984);
        J.\,F.~Donoghue, E.~Golowich and B.\,R.~Holstein, {\it Dynamics
        of the Standard Model} (Cambridge University Press, 1992).
\bibitem{edwin}
        E.~Laermann, Nucl. Phys. A {\bf 610}, 1c (1996);
        E.~Laermann, Nucl. Phys. B (Proc. Suppl.) {\bf 60A}, 180 (1998).
\bibitem{gerry}
        J.~H\"ufner, S.\,P.~Klevansky and P.~Rehberg, Nucl. Phys. A {\bf
        606}, 260 (1996).
\bibitem{qmd}
        J.~Aichelin, Phys. Rep. {\bf 202}, 233 (1991).
\bibitem{hart}
        See
        Ch.~Hartnack, R.\,K.~Puri, J.~Aichelin, J.~Konopka, S.\,A.~Bass,
        H.~St{\"o}cker, W.~Greiner, Euro. Phys. J. {\bf 1}, 151 (1998)
        and references cited therein.
\bibitem{expand}
        E.~Quack and S.\,P.~Klevansky, Phys. Rev C {\bf 49}, 3283 (1994);
        J.~M\"uller and S.\,P.~Klevansky, Phys. Rev. C {\bf 50}, 410 (1994);
        V.~Dmitra\v{s}inovi\'{c}, H.-J.~Schulze, R.~Tegen and
        R.\,H.~Lemmer, Ann. Phys. (NY) {\bf 238}, 332 (1995).
\bibitem{elaste}
        P.~Zhuang, J.~H\"ufner, S.\,P.~Klevansky and L.~Neise, Phys. Rev. D
        {\bf 51}, 3728 (1995);
        P.~Rehberg, S.\,P.~Klevansky and J.~H\"ufner, Nucl. Phys. A {\bf
        608}, 356 (1996).
\bibitem{su2hadron}
        J.~H\"ufner, S.\,P.~Klevansky, E.~Quack and P.~Zhuang, Phys.
        Lett. B {\bf 337}, 30 (1994).
\bibitem{su3hadron}
        P.~Rehberg, S.\,P.~Klevansky and J.~H\"ufner, Phys. Rev. C {\bf
        53}, 410 (1996).
\end{references}
\end{document}